\documentclass{PoS}
\usepackage{graphicx}
\usepackage{tabularx}
\usepackage{multicol}

\newcommand{\figscale}{0.55}
\newcommand{\Npoly}{N_{\mathrm{poly}}}
\renewcommand{\tabcolsep}{4.5pt}

\rightline{\sffamily UTCCS-P-23, HUPD-0605}\vspace*{-10mm}

\title{An application of the UV-filtering preconditioner to the Polynomial Hybrid Monte Carlo algorithm}

\ShortTitle{An application of the UV-filtering preconditioner ...}

\author{PACS-CS Collaboration:}
\author{
\speaker{K-I.~Ishikawa}$^{a}$\thanks{Email: ishikawa@theo.phys.sci.hiroshima-u.ac.jp},
S.~Aoki$^{b,c}$,
T.~Ishikawa$^{d}$,
N.~Ishizuka$^{b,d}$,
K.~Kanaya$^{b}$,
Y.~Kuramashi$^{b,d}$,
M.~Okawa$^{a}$,
Y.~Taniguchi$^{b,d}$,
A.~Ukawa$^{b,d}$,
T.~Yoshi\'{e}$^{b,d}$\\
  \llap{$^a$}Department of Physics, Hiroshima University, Higashi-Hiroshima 739-8526, Japan\\
  \llap{$^b$}Graduate School of Pure and Applied Sciences, University of Tsukuba, Tsukuba 305-8571, Japan\hspace*{-1em}\\
  \llap{$^c$}Riken BNL Research Center, Brookhaven National Laboratory, Upton, NY 11973, USA\\
  \llap{$^d$}Center for Computational Physics, University of Tsukuba, Tsukuba 305-8577, Japan
}

\abstract{
We apply the UV-filtering preconditioner, previously used to improve the 
Multi-Boson algorithm, to the Polynomial Hybrid Monte Carlo (UV-PHMC) algorithm. 
The performance test for the algorithm is given  
for the plaquette gauge action and the $O(a)$-improved Wilson action 
at $\beta=5.2, c_{\mathrm{sw}}=2.02, M_{\pi}/M_{\rho}\sim 0.8$ and $0.7$ 
on a $16^3\times 48$ lattice.
We find that the UV-filtering reduces the magnitude of the molecular dynamics force from the
pseudo fermion by a factor 3 by tuning the UV-filter parameter.
Combining with the multi-time scale molecular dynamics integrator we achieve a factor 2 improvement.}

\FullConference{XXIV International Symposium on Lattice Field Theory\\
                 July 23-28 2006\\
                 Tucson Arizona, US}

\begin{document}

\section{Introduction}
\vspace*{-0.5em}
Recent progress of lattice QCD simulations with dynamical flavors relies on the development of
numerical algorithms and computational facilities.
While the computational power has increased to a multi-Tera-flops level, 
we cannot yet simulate QCD at realistic quark masses.
To overcome this status various numerical algorithms for dynamical lattice QCD
have been proposed. 

The standard algorithm to simulate dynamical lattice QCD is the Hybrid Monte Carlo (HMC)
algorithm~\cite{Duane:1987de}. 
Hence most of recent improvements on the lattice QCD algorithm aim to speed up the HMC algorithm.
There are two key technologies in the literature;
\begin{enumerate}
\setlength{\itemsep}{-1pt}
\item[(1)] Decouple UV and IR fermionic modes by preconditioning lattice Dirac operator, and modify
    the HMC Hamiltonian~\cite{deForcrand:1996ck}.
\item[(2)] Use the Sexton-Weingarten molecular dynamics (MD) integrator with 
   multi (fictitious) time scales~\cite{Sexton:1992nu},
  in which the IR modes of pseudo-fermions are assigned to the coarser time scales and
  the UV modes to the finer time scales.
\end{enumerate}
Various types of preconditioner have been proposed;
the even/odd site preconditioner~\cite{Gupta:1989kxDeGrand:1990dk,Luo:1996tx,Jansen:1996yt},
Hasenbusch's heavy mass preconditioner~\cite{Hasenbusch:2001ne,AliKhan:2003br,Urbach:2005ji},
L\"{u}scher's even/odd domain decomposition (SAP) preconditioner~\cite{Luscher:2005rx}, 
Polynomial preconditioner~\cite{Kamleh:2005wg}, 
$n$-th root multiple pseudo-fermion trick~\cite{Clark:2006fx}, etc.
These preconditioners combined with the Sexton-Weingarten
MD integrator achieve a remarkable (a factor two to ten ) 
speed up over the naive HMC algorithm.

In this article we investigate the UV-filtering preconditioner~\cite{Alexandrou:1999ii} for 
the $O(a)$-improved Wilson-Dirac~\cite{Sheikholeslami:1985ij} fermions.
The UV-filtering preconditioner~\cite{Alexandrou:1999ii} has been proposed 
for the Multi-Boson (MB) algorithm~\cite{Luscher:1993xx,Galli:1996mx}. 
With this preconditioner the number of external multi-boson 
fields can be significantly reduced, leading to a sizable speed up of 
the MB algorithm~\cite{Alexandrou:1999ii}.
We apply this UV-filtering preconditioner to 
the Polynomial Hybrid Monte Carlo (PHMC) algorithm~\cite{deForcrand:1996ck,Frezzotti:1997ym}.
In the next section we describe our UV-filtered Polynomial Hybrid Monte Carlo (UV-PHMC) algorithm.
The numerical results are presented in Section~\ref{sec:Results},
where we investigate the efficiency of the algorithm for the plaquette gluon action 
and the $O(a)$-improved Wilson quark action with 
$\beta=5.2$, $N_f=2$, $c_{\mathrm{sw}}=2.02$, $M_{\pi}/M_{\rho}\sim 0.8$ and $0.7$ on a 
$16^3\times 48$ lattices.

Using the PACS-CS computer~\cite{Aoki:2005sw} the PACS-CS collaboration is planning to 
further promote the $N_f=2+1$ lattice QCD project that has been started by the CP-PACS/JLQCD
joint collaboration~\cite{Ishikawa:2005mj}.
The UV-filtered PHMC (UV-PHMC) algorithm will be applied to the single flavor part of the $N_f=2+1$ 
simulations. A status report of the PACS-CS collaboration is given in Ref.~\cite{Kuramasi}.

\vspace*{-0.5em}
\section{Algorithm}
\label{sec:Algorithm}
\vspace*{-0.5em}
To describe the UV-PHMC algorithm we start with the lattice QCD partition function with the $O(a)$-improved Wilson
fermion in the symmetrically even/odd-site preconditioned form~\cite{Jansen:1996yt,Aoki:2001pt}.

\begin{eqnarray}
  {\cal Z}&=&\int{\cal D}U \det[D[U]]^{N_f} e^{-S_{G}[U] - S_{\mathrm{clv}}[U]},
  \label{eq:QCDpart} \\
  S_{\mathrm{clv}}[U]&=&-N_f \mathrm{Tr}[\mathrm{Log}[T^{-1}]],\label{eq:Sclv}\\
  D &=& 1_{ee} - T_{ee} M_{eo} T_{oo} M_{oe}  = 1_{ee} - \hat{M}_{ee},\\
  T &=& (1 + c_{\mathrm{sw}} \kappa \sigma\cdot F)^{-1}.
\end{eqnarray}
where $U$ denotes gauge links, $S_G[U]$ is a gauge action, $T$ is the local clover term with
the clover leaf field strength $F$ and the clover coefficient $c_{\mathrm{sw}}$. 
$M_{eo}$ ($M_{oe}$) is the single hopping matrix jumping from odd (even) sites to even (odd) sites, 
and $D $ and $\hat{M}_{ee}$ operate only on even sites. 
We apply the UV-filter preconditioner to $D$ in Eq.~(\ref{eq:QCDpart}).

\subsection{The UV-filter}
 We introduce the UV-filter preconditioner $P[U]$ as
 \begin{equation}
  \label{eq:UVfilter}
   P[U]=\exp[ s \hat{M}_{ee}],
 \end{equation}
where `$s$' is a tunable parameter (UV-filter parameter). 
We can understand that this operator is a preconditioner by setting $s=1$ as follows.
\begin{equation}
 Q[U]\equiv P[U]D[U] = \exp[\hat{M}_{ee}](1-\hat{M}_{ee}) = 1 - \frac{(\hat{M}_{ee})^2}{2}- \frac{(\hat{M}_{ee})^3}{3} - \ldots.
\end{equation}
Since $\hat{M}_{ee}$ is $O(\kappa^2)$, $P[U]$ removes $O(\kappa^2)$ term and
the preconditioned operator $Q$ becomes $1 + O(\kappa^4)$ close to identity matrix~\cite{Alexandrou:1999ii}.
Using this preconditioner one can rewrite the quark determinant as
\begin{eqnarray}
  \det[D[U]]^{N_f}&=&\det[(P[U])^{-1}P[U]D[U]]^{N_f}\nonumber \\
                  &=&\det[Q[U]]^{N_f}\exp[- s N_f\mathrm{Tr}[\hat{M}_{ee}]] 
               \equiv\det[Q[U]]^{N_f}\exp[- N_f S_{\mathrm{uv}[U]}],
\end{eqnarray}
where
\begin{equation}
  S_{\mathrm{uv}}[U] = s \mathrm{Tr}[\hat{M}_{ee}]
= s \kappa^2 \sum_{n,\mu}\mathrm{tr}_{\mathrm{color,dirac}}[
              T(n)(1-\gamma_\mu)U_{\mu}(n)T(n+\mu)(1+\gamma_\mu)U^{\dag}_{\mu}(n)].
\label{eq:Suv} 
\end{equation}
Note that $S_{\mathrm{uv}}$ is still a local action and vanishes when $c_{\mathrm{sw}}=0$. 
For the unimproved Wilson fermion further preconditioning, which removes $O(\kappa^4)$ term, has 
been investigated in the MB algorithm~\cite{Alexandrou:1999ii}. In our improved case 
$O(\kappa^4)$ preconditioning results in a complicated (non ultra-local) action for $S_{\mathrm{uv}}$
and we do not investigate the $O(\kappa^4)$ preconditioner in this article.
We employ Eq.~(\ref{eq:UVfilter}) for the UV-filter. 

\subsection{Hybrid Monte Carlo algorithm with the UV-filter}
By applying the polynomial approximation $P_{\Npoly} \sim (Q[U])^{-1}$, and introducing
a pseudo-fermion $\phi$ and a fictitious momenta $\Pi$ for gauge links, we obtain
\begin{eqnarray}
\label{eq:UVPHMCaction}
  {\cal Z}&=&\int{\cal D}\Pi{\cal D}U{\cal D}\phi^{\dag}{\cal D}\phi
\det[W[U]]^{N_f}e^{-H[U,\phi^{\dag},\phi]},\\
H[U,\phi^{\dag},\phi]&=&
\mathrm{Tr}[\Pi^2] + S_{G}[U] +S_{\mathrm{clv}}[U] + N_f S_{\mathrm{uv}}[U] + S_{Q}[U,\phi^{\dag},\phi],\\
S_Q &=&| P_{\Npoly}[\hat{M}_{ee}] \phi|^2,
\label{eq:SQ}\\
W   &=& P_{\Npoly}[\hat{M}_{ee}](1-\hat{M}_{ee})\exp[s \hat{M}_{ee}],\\
P_{\Npoly}[\hat{M}_{ee}]&=& \sum_{k=0}^{\Npoly} c_{k}(\hat{M}_{ee})^{k},
\label{eq:poly}
\end{eqnarray}
where $s$ and $c_{k}$ are tuned to satisfy $W \sim 1$. 
The choice of $s$ and $c_k$ will be described in the next subsection.
When $s=0$ this action reduces to that for 
the normal PHMC algorithm. Eq.~(\ref{eq:poly}) applies to the $N_f=2$ case. 
For the $N_f=1$ case, we use the factorized polynomial instead of Eq.~(\ref{eq:poly})
as described in Ref.~\cite{Aoki:2001pt}.
The actions $S_{G}$, $S_{\mathrm{clv}}$, and  $S_{\mathrm{uv}}$ can be classified in the UV part, 
and $S_{Q}$ in the IR part.
The effect of $W$ is incorporated by the noisy-Metropolis test as having been used 
in the MB algorithm~\cite{Galli:1996mx} and CP-PACS/JLQCD's previous studies~\cite{Ishikawa:2005mj}.
We investigate the PHMC algorithm with Eq.~(\ref{eq:UVPHMCaction}).

The flow of the algorithm is almost the same as that described in Ref.~\cite{Aoki:2001pt} except for
the following minor changes. 
\begin{itemize}
\setlength{\itemsep}{-3pt}
\item For the MD integrator we employ the Sexton-Weingarten MD integrator~\cite{Sexton:1992nu} 
      with two time-step scales (UV and IR).
   The PUP-order integration scheme has been used in the literature.
   In this study we test the following UPU-order integrator;
      \begin{equation} {\hspace*{-0.5em}
        \left[
             \left\{
               U\left(\frac{\delta\tau_0}{2}\right)
               P_{\mathrm{UV}}\left(\delta\tau_0\right)
               U\left(\frac{\delta\tau_0}{2}\right)
             \right\}^{\frac{N_0}{2}}
             P_{\mathrm{IR}}\left(\delta\tau_1\right)
             \left\{
               U\left(\frac{\delta\tau_0}{2}\right)
               P_{\mathrm{UV}}\left(\delta\tau_0\right)
               U\left(\frac{\delta\tau_0}{2}\right)
             \right\}^{\frac{N_0}{2}}
        \right]^{N_1},}
        \label{eq:UPUMD}
      \end{equation}
   where $\delta\tau_0=\tau/N_0/N_1$, $\delta\tau_1=\tau/N_1$, and $\tau$ is trajectory length.
   $U(\delta \tau)$ integrates gauge links by $\delta\tau$, 
   $P_{X}(\delta \tau)$ integrates gauge momenta by $\delta\tau$.
   The UV-modes are integrated by $P_{\mathrm{UV}}$ and the IR-modes by
   $P_{\mathrm{IR}}$. $N_0$ and $N_1$ are the number of time-steps in each trajectory and $N_0$ 
   should be an even number in this scheme.  

\item For the single flavor case we need to take the square root of the correction matrix $W$ to do the noisy-Metropolis test.
We tested a new algorithm of Ref.~\cite{ResKryMatfunc} for the matrix square root problem. The
algorithm utilizes the Krylov-subspace method via the Arnoldi factorization.
Since the use of the Krylov-subspace method does not significantly affect the whole efficiency of the UV-PHMC algorithm,
we will skip the details of the matrix square root algorithm in this article.

\item For the UV-filter we need to calculate the matrix exponential of $\hat{M}_{ee}$.
We tested the following three methods;
(i) the Taylor expansion approximation method,
(ii) the Pad\'{e} approximation method (without multi-shift solver),
(iii) the Krylov-subspace approximation method~\cite{expokit}. 
We employ the Taylor expansion approximation method (i) 
because of its simplicity and moderate efficiency.
The truncation error of the Taylor approximation is controlled by monitoring the spectrum norm of $\hat{M}_{ee}$.
\end{itemize}

\subsection{Choice of polynomial coefficients}
In order to minimize the cost of the algorithm, the UV-filter coefficient $s$ and the polynomial 
coefficients $c_k$ should be chosen to satisfy the condition that $W \sim 1$ with a small $\Npoly$. 
We investigated the following two coefficient schemes.
\begin{enumerate}
\item[(A)] Taylor expansion method: By expanding $[(1-\hat{M}_{ee})\exp[s\hat{M}_{ee}]]^{-1}$ with respect to $\hat{M}_{ee}$,
  we obtain
  \begin{equation}
    [(1-\hat{M}_{ee})\exp[s\hat{M}_{ee}]]^{-1} \sim \sum_{k=0}^{\Npoly} c_k (\hat{M}_{ee})^{k},
    \ \ \ \mbox{with}\ \ \  c_k = \sum_{j=0}^{k}\frac{(-s)^j}{j!}. \label{eq:taylorcoef}
  \end{equation}
  This is nothing but the hopping matrix expansion.
\item[(B)] Adopted polynomial method~\cite{Alexandrou:1999ii}: This method minimizes the following function;
    \begin{equation}
      R(\vec{c},s)=\left| \left(\left[\sum_{k=0}^{\Npoly}c_k(\hat{M}_{ee})^k\right](1-\hat{M}_{ee})\exp[s\hat{M}_{ee}] 
                                       - 1\right)\eta\right|^2 = |(W-1) \eta|^2,
    \end{equation}
    where $\eta$ is a Gaussian random noise vector and $\vec{c}=(c_0, c_1, c_2, \ldots, c_{\Npoly})$. This method has been used
    in the MB algorithm as described in Ref.~\cite{Alexandrou:1999ii}. 
    The function minimization is carried out by a linear fitting for $\vec{c}$ with a fixed $s$ 
    followed by Newton's method for $s$. We take several thermalized gauge configurations for the fitting.
\end{enumerate}

\FIGURE{
  \centering
  \includegraphics[scale=\figscale,clip]{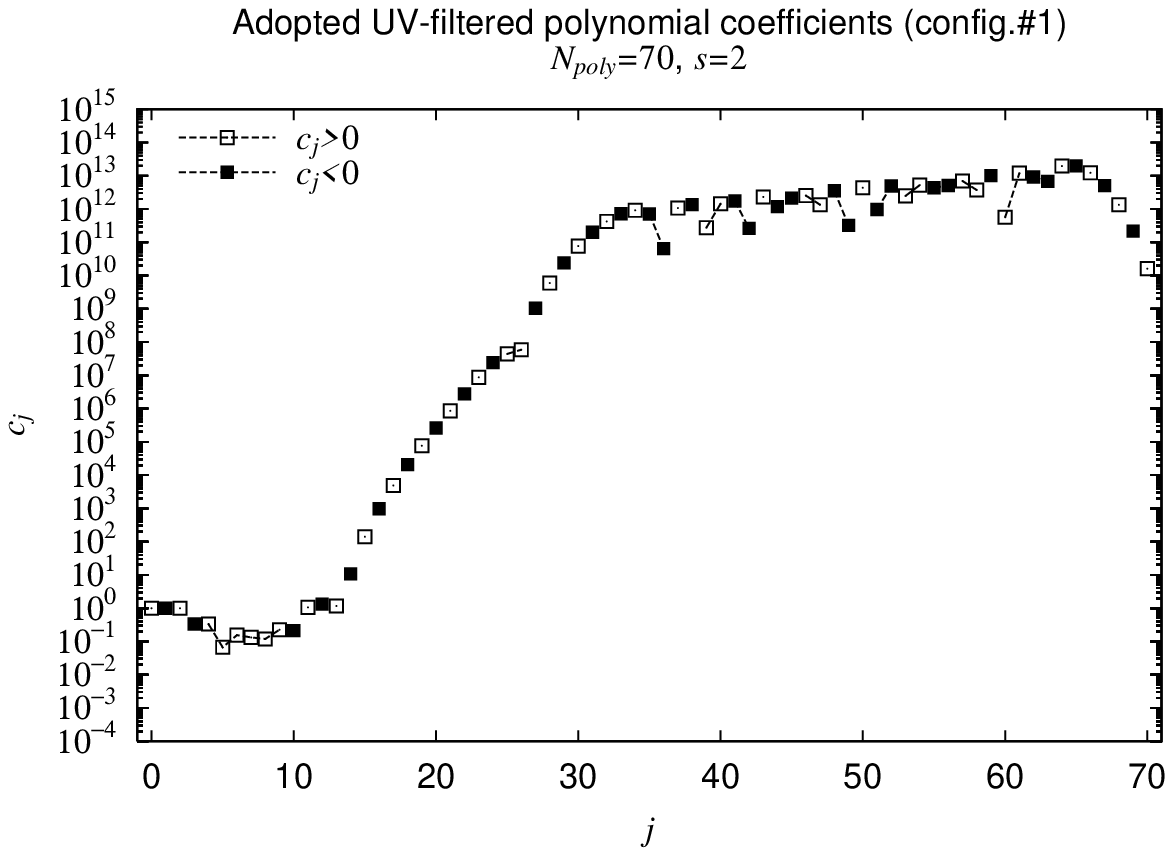}
  \vspace*{-1em}
  \caption{Adopted polynomial coefficients with
           $s=2$ and $\Npoly=70$ determined on a 
           thermalized configuration at $16^3\times 48$, 
           $\beta=5.2$, $N_f=2$, $\kappa=0.1350$, $c_{\mathrm{sw}}=2.02$. 
           Similar behavior is observed for different $s$ and $\Npoly$. 
          The configuration dependence is negligible.}
 \label{fig:polycoef}
}

Figure~\ref{fig:polycoef} shows the polynomial coefficients with the Adopted polynomial method (B).
We observed that the dynamic range of the coefficients spreads from $O(10^{-2})$ to $O(10^{13})$.
This means that careful treatment of the numerical accuracy and stability is required
to compute the polynomial $P_{\Npoly}$ within a finite precision arithmetic.
Although we used double precision arithmetic and Clenshaw recurrence formula to construct the polynomial,
we could not maintain good accuracy and stability for $P_{\Npoly}$.
In the rest of paper, therefore, we employ the Taylor expansion method (A) and Eq.~(\ref{eq:taylorcoef}) 
for the coefficients.

\vspace*{-0.5em}
\section{Numerical Results}
\label{sec:Results}
\vspace*{-0.5em}
\TABLE{
  \begin{tabular}[t]{cccc}\hline
    Simulation & H & L \\\hline
    $\kappa$   & 0.1340 & 0.1350 \\\hline
    $M_{\mathrm{PS}}/M_{\mathrm{V}}$   & $\sim$0.8 & $\sim$0.7 \\\hline
  \end{tabular}
  \caption{Simulation parameters.}
  \label{tab:para}
}

We employ the plaquette Wilson gauge action for $S_{G}$.
Two quark masses are studied
at $\beta=5.2$ on a $16^3\times 48$ lattice for $N_f=2$ and $c_{\mathrm{sw}}=2.02$ 
(see Table~\ref{tab:para}).
The simulations are denoted as H (heavy quark mass)  and L (lighter quark mass).
Table~\ref{tab:Hresult} and \ref{tab:Lresult} present the simulation results for 
the norm of MD force for each sector and simulation statistics. 
The trajectory length $\tau$ is set to unity, and $N_0$ ($N_1$) is the number of time steps in
for UV scale (IR scale) in a single trajectory.
For comparison we tabulate the PHMC algorithm and the symmetrically even/odd-site 
preconditioned HMC (SHMC) in the tables.  $s=\mathrm{PHMC}$ is equivalent to $s=0$ of UV-PHMC.
The definition of the force norm is 
$|F|=\sum_{n,\mu}\mathrm{Tr}[F_{\mu}(n)F^{\dag}_{\mu}(n)]/2/L^3/T$, where
$L=16$, $T=48$, and $F_{\mu}(n)$ is the MD force to drive gauge link $U_{\mu}(n)$.
$P_{\mathrm{HMC}}$ is the HMC Metropolis test acceptance rate, and $P_{\mathrm{GMP}}$  
the global noisy Metropolis test acceptance rate. We monitored 
the averaged number of matrix vector multiplication of $\hat{M}_{ee}$ to move forward the algorithm 
by one trajectory (``Mult/traj'' in the tables).

Without UV-filtering the MD force from pseudo-fermion (gauge) action $|F_Q|$ ($|F_G|$) is about 1.4 (4.5).
The contribution from $|F_{\mathrm{uv}}|$ and $|F_{\mathrm{clv}}|$ is smaller than 
that from $|F_{\mathrm{Q}}|$ and $|F_{\mathrm{G}}|$.
We observe that $|F_Q|$ depends on the UV-filter parameter $s$ and takes its minimum value at $s=1$.
The reduction of $|F_Q|$ from $s=0$ to $s=1$ is about a factor three for both (L) and (H) lattices.
Exploring $N_0$ and $N_1$ by keeping the HMC acceptance rate $P_{\mathrm{HMC}}$ around $0.7$, we get
the computational cost reduction in Mult/traj by a factor two at $s=1$ for both (L) and (H) lattices.
Comparing the efficiency between the PUP-order and the UPU-order schemes at the (L) parameter, 
a small gain in the HMC acceptance is observed for the UPU-order scheme.

Table~\ref{tab:LresultNf11} shows the result with $N_f=1+1$ simulation. 
The action contains two pseudo-fermions where one pseudo-fermion represents single flavor.
The force norm $|F_Q|$ contains the force from both pseudo-fermions. 
As observed in Ref.~\cite{Aoki:2001pt} the HMC acceptance rate becomes better than that 
with the $N_f=2$ single pseudo-fermion simulation. 
The reason of the improvement using multiple pseudo-fermion is explained in Refs.~\cite{Clark:2006fx,Kennedy:2006ax}.

\begin{table}[t]
  \centering
{\scriptsize
  \begin{tabular}{ccc|cccc|ccc}\hline
   $[\tau,N_1,N_0]$ &
     $s$ &   traj. & $|F_{\mathrm{clv}}|$ & $|F_{\mathrm{G}}|$ & $|F_{\mathrm{Q}}|$ & $|F_{\mathrm{uv}}|$ 
   & $P_{\mathrm{HMC}}$ & $P_{\mathrm{GMP}}$ & Mult/traj \\\hline
   $[1,20,6]$ &  1.1 & 1500& 0.282888(14) & 4.51815(22) & 0.493642(97)& 0.0455531(74)&0.742(14)&0.894(10)& 5587(13)\\
   $[1,25,4]$ &  1.1 & 1000& 0.28329(40)  & 4.504(16)   & 0.49404(89) & 0.04577(24)  &0.847(19)&0.868(13)& 6679(17)\\
   $[1,56,0]$ &
        {\tiny PHMC} & 800 & 0.282814(16) & 4.52032(34) & 1.33903(28) & -            &0.780(24)&0.895(17)&12456(15)\\\hline
  \end{tabular}
}\vspace*{-0.2em}
  \caption{Simulation statistics with $\Npoly=80$ (H) parameter.}
  \label{tab:Hresult}
\end{table}

\begin{table}[t]
  \centering
{\scriptsize
  \begin{tabular}{ccc|cccc|ccc}\hline
   $[\tau,N_1,N_0]$ &
     $s$ &   traj. & $|F_{\mathrm{clv}}|$ & $|F_{\mathrm{G}}|$ & $|F_{\mathrm{Q}}|$ & $|F_{\mathrm{uv}}|$ 
   & $P_{\mathrm{HMC}}$ & $P_{\mathrm{GMP}}$ & Mult/traj \\\hline
   $[1,25,8]$ &  0.0 &  10 & 0.286702(53) & 4.53781(27) & 1.490(25)   & -            & -       & - & - \\
   $[1,25,8]$ &  0.5 &  10 & 0.286761(40) & 4.53801(23) & 0.8285(16)  & 0.0211877(54)& -       & - & - \\
   $[1,25,8]$ &  1.0 &  10 & 0.286812(50) & 4.53933(43) & 0.52395(64) & 0.042375(16) & -       & - & - \\
   $[1,25,8]$ &  1.1 & 1000& 0.2867765(90)& 4.53843(17) & 0.52682(26) & 0.0466221(59)&0.669(17)&0.863(16)& 12607(26)\\
   $[1,25,4]$ &  1.1 & 1000& 0.286783(11) & 4.53845(30) & 0.52650(23) & 0.0466247(85)&0.692(18)&0.868(16)& 12640(29)\\
   $[1,25,2]$ &  1.1 & 1000& 0.286798(11) & 4.53813(18) & 0.52763(29) & 0.0466381(66)&0.664(17)&0.902(11)& 12596(29)\\
   $[1,25,0]$ &  1.1 & 1200& 0.286841(16) & 4.53814(36) & 0.52693(41) & 0.0466632(92)&0.274(29)&0.882(21)& 11992(45)\\
   $[1,25,4]^{*}$ &  1.1 &  900& 0.286805(22) & 4.53784(63) & 0.52736(29) & 0.046644(20) &0.636(13)&0.891(14)& 12958(35)\\
   $[1,25,2]^{*}$ &  1.1 & 1000& 0.286790(13) & 4.53858(34) & 0.52609(43) & 0.046627(10) &0.531(26)&0.864(17)& 12816(45)\\
   $[1,25,8]$ &  1.5 &  10 & 0.286751(28) & 4.53789(24) & 0.74044(39) & 0.063577(17) & -       & - & - \\
   $[1,25,8]$ &  2.0 &  10 & 0.286861(33) & 4.53787(27) & 1.25248(53) & 0.084818(31) & -       & - & - \\
   $[1,70,0]$ &
        {\tiny PHMC} & 1100& 0.286793(13) & 4.53816(29) & 1.39445(57) & -            &0.762(16)&0.871(14)& 30385(20)\\
   $[1,70,0]$ & 
        {\tiny SHMC} & 340 & 0.286771(16) & 4.53909(21) & 1.39161(48) & -            & 0.80(3) & - & 37491(166) \\\hline
  \end{tabular}
}\vspace*{-1em}
  \caption{Simulation statistics with $\Npoly=160$ (L) parameter. 
          [{\footnotesize $^{*}$: the PUP-order MD integrator is used.}]}
  \label{tab:Lresult}
\end{table}

\newcolumntype{C}{>{\centering\arraybackslash}X}
\def\tabularxcolumn#1{m{#1}}
\begin{table}[t]
  \centering
{\scriptsize
  \begin{tabularx}{\hsize}{ccc|cccc|cCc}\hline
   $[\tau,N_1,N_0]$ &
     $s$ &   traj. & $|F_{\mathrm{clv}}|$ & $|F_{\mathrm{G}}|$ & $|F_{\mathrm{Q}}|$ & $|F_{\mathrm{uv}}|$ 
   & $P_{\mathrm{HMC}}$ & $P_{\mathrm{GMP}}$ & Mult/traj \\\hline
   $[1,25,4]$ &  1.1 & 1000 &0.2867632(93)&4.53873(20)&0.363581(73)&0.0466130(72)
              & 0.906(12) 
              & 0.937(10) 0.943(11)
              & 18896(110)
 \\\hline
  \end{tabularx}
}\vspace*{-1em}
  \caption{Simulation statistics with $N_f=1+1$, $\Npoly=160$ (L) parameter. }
  \label{tab:LresultNf11}
\end{table}

\vspace*{-0.5em}
\section{Summary and outlook}
\label{sec:Summary}
\vspace*{-0.5em}
In this work we have presented the effectiveness of the
UV-filtering preconditioner to the Polynomial Hybrid Monte Carlo algorithm.
The simulations have been carried out on lattices with moderate size and moderate quark masses.
The UV-filtering preconditioner reduces the magnitude of MD force of
the pseudo-fermion part and enables us to extend the MD time-step size of the pseudo-fermion.
The gain in computational cost is a factor two on the lattices we have investigated.
We have also tested the $N_f=1+1$ case to confirm the efficiency of the single flavor algorithm.
The UV-filtering for the Wilson type fermions is applicable to the heavy mass preconditioner by Hasenbusch~\cite{Hasenbusch:2001ne}
and the polynomial filtering~\cite{Kamleh:2005wg} and further speed up is expected.
We are planning to apply the UV-filtered PHMC algorithm to the single flavor part 
of $N_f=2+1$ simulations.

\ \\
\indent
The simulation has been carried out on Hitachi SR11000 at 
Information Media Center of Hiroshima University.
This work is supported in part by the Grant-in-Aid of the Ministry of Education 
(Nos. 13135204, 13135216, 15540251, 16540228, 16740147, 17340066, 17540259, 18104005, 18540250, 18740130).

\end{document}